\pdfoutput=1

\documentclass[journal,10pt]{IEEEtran}
\usepackage{relsize}
\usepackage{amsmath}
\usepackage{multicol}
\usepackage{amsfonts}
\usepackage{amsmath}
\usepackage{graphicx}
\usepackage{textcomp}
\usepackage{float}
\usepackage{epstopdf}
\usepackage{mathtools}
\usepackage{multicol}
\usepackage{float}
\usepackage{lipsum}

\input{tcilatex}
\begin{document}

\title{Analysis of Average Consensus Algorithm for Asymmetric Regular Networks}
\author{Sateeshkrishna Dhuli and Y. N. Singh~%
\IEEEmembership{Senior Member,~IEEE} 
}
\maketitle

\begin{abstract}
Average consensus algorithms compute the global average of sensor data in a distributed fashion using local sensor nodes. Simple execution, decentralized philosophy make these algorithms suitable for WSN scenarios. Most of the researchers have studied the average consensus algorithms by modeling the network as an undirected graph. But, WSNs in practice consist of asymmetric links and the undirected graph cannot model the asymmetric links. Therefore, these studies fail to study the actual performance of consensus algorithms on WSNs. In this paper, we model the WSN as a directed graph and derive the explicit formulas of the ring, torus, $r$-nearest neighbor ring, and $m$-dimensional torus networks. Numerical results subsequently demonstrate the accuracy of directed graph modeling. Further, we study the effect of asymmetric links, the number of nodes, network dimension, and node overhead on the convergence rate of average consensus algorithms.
\end{abstract}

{} 

\begin{IEEEkeywords}
Average Consensus Algorithms, Regular Graphs, Asymmetric Links, Directed Graph, Convergence Rate, Wireless Sensor Networks.
\end{IEEEkeywords}

\IEEEpeerreviewmaketitle

\section{Introduction}
\IEEEPARstart{C}{onsensus} algorithms have been widely studied in the literature due to their decentralized philosophy and simple execution \cite{olfati2007}, \cite{olshevsky2009}, \cite{xiao2007}. These algorithms can be utilized when global network topology information is not known, and the network consists of power constrained nodes. In contrast to centralized algorithms, these algorithms compute the desired statistics at every node without the need of any fusion center. Hence, these algorithms are quite suitable for WSN scenarios. Consensus algorithms are iterative in nature, and their performance is measured by convergence rate\cite{olshevsky2011}, \cite{cao2005}. Convergence rate of the consensus algorithms have been widely studied in the literature, most of the prior works have modeled the networks as an undirected graph due to the computational tractability. However, undirected graphs cannot model the applications which involve asymmetric links and may not characterize the actual network’s performance. In practice, wireless channels in low power wireless networks such as WSNs are known to be time-varying, unreliable, and asymmetric \cite{zhou2006}, \cite{zamalloa2007}, \cite{yang2006}, \cite{kotz2003}, \cite{ganesan2002}, \cite{li2013}. Therefore, it is important to consider the WSN as a directed graph to accurately estimate the convergence rate. Convergence rate is characterized by the graph Laplacian eigenvalues\cite{asadi2017}, \cite{schug2014}. Estimating the convergence rate for large-scale networks is a computationally challenging task. To evaluate the convergence rate, there are many algorithms available in the literature, such as \textit{best constant} weights algorithm, \textit{metropolis–-hastings} weights algorithm, \textit{max-degree} weights algorithm \cite{xiao2007}. In this paper, we employ the \textit{best constant} weights algorithm to derive the explicit expressions of the convergence rate. Recently, WSN has been modeled as an $r$-nearest neighbor network and explicit expressions of convergence time for average consensus algorithms have been derived in \cite{dhuli2015}. However, they considered the undirected graph modeling which cannot study the time-varying wireless channels of WSNs. In \cite{hao}, authors modeled the WSN as a directed graph and proved that asymmetric weights improve the convergence rate of average consensus algorithms. The expected convergence rate of an asymmetric network has been examined in \cite{asadi2017}. In our work, we model the WSN as a directed graph and derive the explicit expressions of convergence rate for regular graphs.

Regular graph models are simple structures which allow the theoretical analysis that incorporates important parameters like connectivity, scalability, network size, node overhead, and network dimension \cite{vanka2010}, \cite{lattice}, \cite{dhuli2015}. These models represent the geographical proximity in the practical wireless sensor networks. In this paper, we model the WSN as a ring, torus, $r$-nearest neighbor network, $m$-dimensional torus networks and derive the explicit expressions for convergence rate of average consensus algorithms. The nearest neighbors can model node's transmission radius or node overhead. We measure the absolute error to investigate the deviation of convergence rate results in directed graph modeling over undirected graph modeling. Our analytic expressions are extremely helpful in designing the optimization frameworks for controlling the performance of average consensus algorithms on WSNs. Our approach avoids the usage of huge computational resources to compute the convergence rate for large-scale WSNs.

\subsection{Organization}
1)In section II, we give a brief review on consensus algorithms\\
2)In section III, we model the WSN as a ring and derive the explicit expressions of convergence rate in terms of number of nodes and network overhead.\\
(3)In section IV, we model the WSN as a torus network and $m$-dimensional torus network and derive the explicit expressions of convergence rate in terms of number of nodes and network dimension.\\
(4)In section V, we model the WSN as a $r$-neighbor neighbor ring network and derive the explicit expressions of convergence rate in terms of nearest neighbors and number of nodes.\\
(5)Finally, in section IV, we present the numerical results and study the effect of network parameters on the convergence rate.
\section{Average Consensus Algorithm}
Let $G= (V,E)$, be a directed graph with node set $V = \left\{ {1,2,......n} \right\}$ and an edge set $E \subseteq V \times V$. Let $A$ denotes the $n\times n$ adjacency matrix of graph $G$, where each entry of $A$ is represented by $a_{ij}$. The degree matrix $D$ is defined as the diagonal matrix whose entry is $d_{ii}$, where $d_{ii}=\sum_{j=1}^{n}a_{ij}=\sum_{j=1}^{n}a_{ji}$. The Laplacian matrix of a graph $G$ is expressed as $L=D-A$, whose entries are
\begin{equation}
l_{ij}  = l_{ji}  = \left\{ \begin{array}{l}
 \deg (v_i )\,\,\,if\,\,j = i ,\\
  - a_{ij} \,\,\,\,\,\,\,\,\,if\,\,j \ne \,i .\\
 \end{array} \right.
 \label{1}
\end{equation}
Let $x_i (0)$ denotes the real scalar variable of node i at $t=0$. Average consensus algorithm computes the average $x_{avg}=\frac{{\sum\nolimits_{i = 1}^n {x_i (0)} }}{n}$ at every node through a decentralized approach which does not require any sink node. At each step, node $i$ carries out its update based on its local state and communication with its direct neighbors. At time instant $t+1$, the real scalar variable at node $i$ is expressed as 
\begin{equation}
x_i (t + 1) = x_i (t) + h\sum\limits_{j \in N_i } {(x_j (t) - x_i (t))} ,\,\,\,i = 1,...,n,
\label{2}
\end{equation}
where `\textit{h}' is a consensus parameter and $N_i$ denotes neighbor set of node `\textit{i}'.
This can be also expressed as a simple linear iteration as 
\begin{equation}
\textbf{x(t + 1)} = W\textbf{x(t)},\,\,\,\,t = 0,1,2...,
\label{3}
\end{equation}
where `\textit{W}' denotes weight matrix, and $W_{ij}$ is a weight associated with the edge $(i,j)$. If we assign equal weight \textit{h} to each link in the network, then optimal weight for a given topology is
\begin{equation}
W_{ij}=\left\{\begin{matrix}
h & if ,\,\,\,\,(i,j) \in E,\\
1-hdeg(\nu_{i})&if,\,\,\,\,i=j,\\
0 & otherwise.
\end{matrix}\right.
\label{4}
\end{equation}
and Weight matrix is given by
\begin{equation}
W = I - hL.
\label{5}
\end{equation}
where `\textit{I}' is a $n\times n$ identity matrix. Let $\lambda _n (W)$ be the $n^{th}$ eigenvalue of $W$, then $\lambda _n (W)=1-h\lambda _n (L)$ satisfies
 \begin{equation}
1=\lambda _{1}(W)> \lambda _{2}(W)> \lambda _{3}(W).........\lambda _{n}(W).
\label{6}
\end{equation}
and let $\lambda _{n}(L)$ be the $n^{th}$ eigenvalue of Laplacian matrix satisfies
\begin{equation}
0=\lambda _{1}(L)< \lambda _{2}(L)< \lambda _{3}(L).........\lambda _{n}(L).
\label{7}
\end{equation}
Convergence rate of a average consensus algorithm can be measured by the spectral gap $\left| {1 - \lambda _2 (L)} \right|$\cite{hao}, \cite{asadi2017}, \cite{schug2014}. In this work, we employ the \textit{best constant weights} algorithm to derive the closed-form expressions of convergence rate. \textit{Best constant weights} algorithm gives the fastest convergence rate among the other uniform weight methods \cite{toulouse2015}, \cite{2011local}.
\subsection{Best Constant  Weights Algorithm}
Derive the generalized eigenvalue expression of Laplacian matrix and follow the below steps. \\ 
(1)Compute the second smallest eigenvalue of Laplacian matrix ($\lambda _1 \left( L \right)$) and largest eigenvalue of Laplacian matrix ($\lambda _{n - 1} \left( L \right)$). \\
(2)Obtain the consensus parameter ($h$) using 
\begin{equation}
{\left| {1 - h\lambda _1 \left( L \right)} \right| = \left| {1 - h\lambda _{n - 1} \left( L \right)} \right|}
\label{8}
\end{equation}\\
(3)Substitute the `\textit{h}'  in $\left| {1 - h\lambda _1 \left( L \right)} \right|$ and obtain the convergence parameter ($\gamma=h\lambda _1 \left( L \right)$). \\
(4)Finally, evaluate the convergence rate ($R$) using 
\begin{equation}
R=1-\gamma.
\label{9}
\end{equation}

 \section{Explicit Formulas of Convergence Rate for Ring Networks}
In this section, we derive the explicit expressions of convergence rate for ring and $r$-nearest neighbor ring networks. Ring network with asymmetric links is as shown in the Fig. 1. We assume that forward link weight is $\frac{1-a}{2}$ and backward link weight is $\frac{1+a}{2}$, here 
`\textit{a}' denotes asymmetric link factor.
\begin{figure}[tbp]
\centering
\includegraphics[totalheight=6cm]{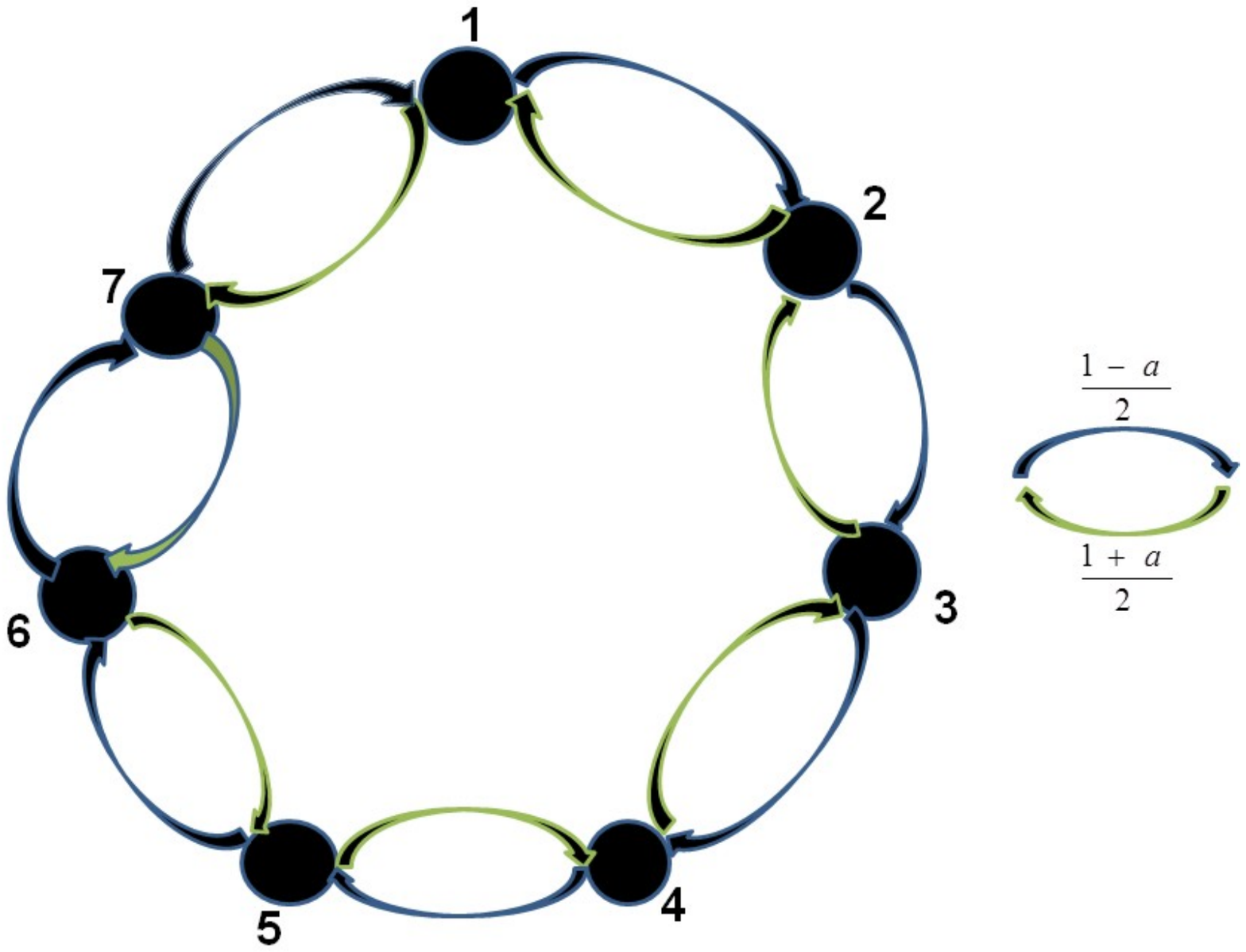}
\caption{Asymmetric Ring Network}
\label{fig:verticalcell}
\end{figure}
\begin{figure}[tbp]
\centering
\includegraphics[totalheight=6cm]{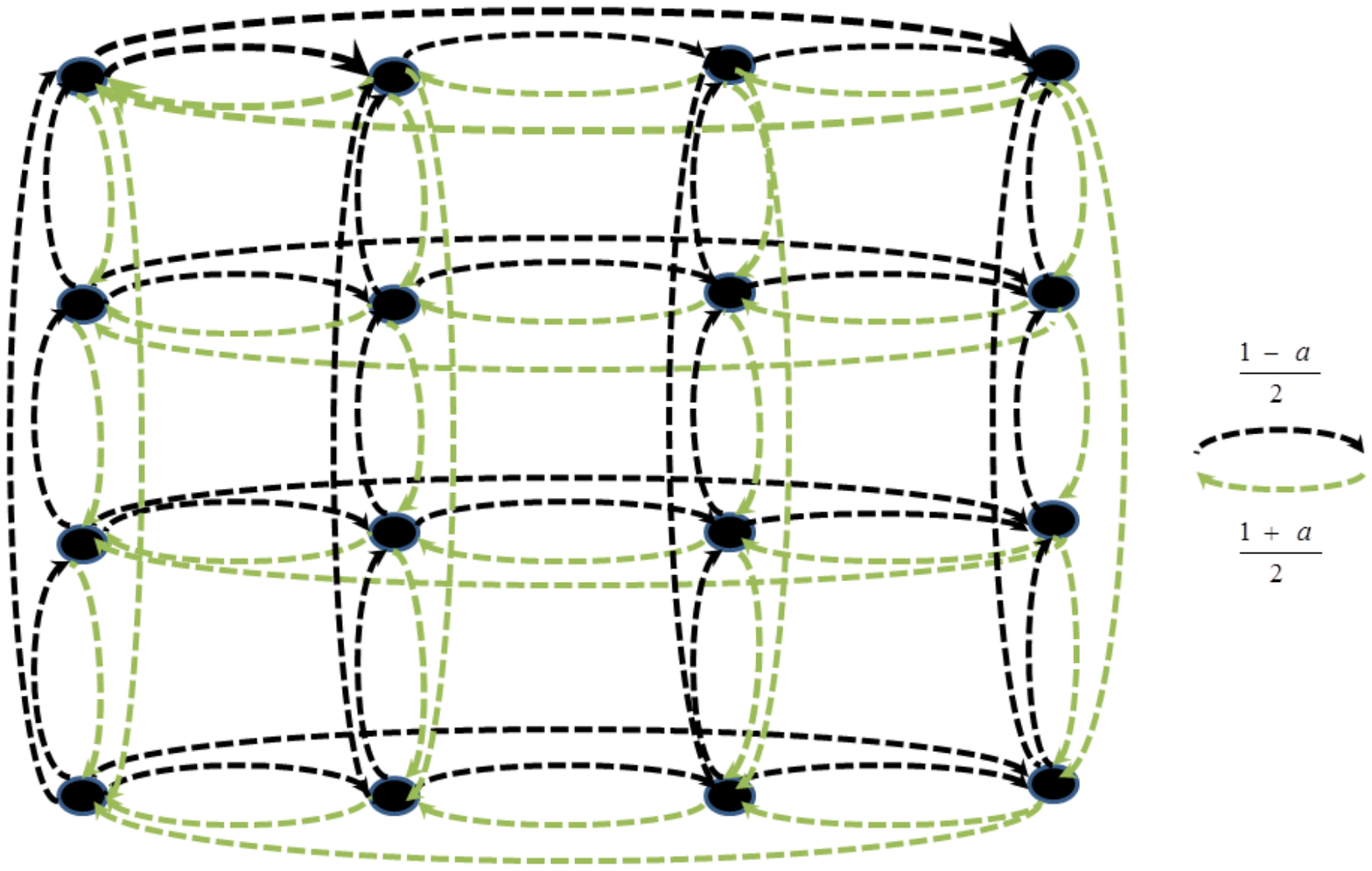}
\caption{Asymmetric Torus Network}
\label{fig:verticalcell}
\end{figure} \\
$Definition$ $1$: The $(j+1)^{th}$ eigenvalue \cite{circulant} of a circulant matrix $circ\{a_1 ,a_2 ......a_n\}$ is defined as
\begin{equation}
\lambda _j  = a_1  + a_2 \omega ^{j}  + .............. + a_n \omega ^{(n - 1)j},
\label{10}
\end{equation}
where $\omega= e^\frac{2\pi i}{n}$ and $\{ a_l \} _{l = 1}^n $ are row entries of circulant matrix.\\
$Theorem$ $1$: The $(j+1)^{th}$ eigenvalue of Laplacian matrix of a ring network for even number of nodes is expressed as
\begin{equation}
\lambda _j (L) = 1 - \cos \frac{{2\pi j}}{n} + ai\sin \frac{{2\pi j}}{n}
\label{11}
\end{equation}
$Proof$ $1$: Asymmetric ring network is as shown in the Fig. 1. Thus, Laplacian matrix can be written as
\begin{equation}
L = circ\{ 1,\frac{{ - 1 + a}}{2},\underbrace {0,0,..,0}_{n - 3\,\,terms},\frac{{ - 1 - a}}{2}\} 
\label{12}
\end{equation}
Using (\ref{10}), $(j+1)^{th}$ eigenvalue of a Laplacian matrix can be written as
\begin{equation}
\lambda_j (L) = 1 - \cos \frac{{2\pi j}}{n} + ai\sin \frac{{2\pi j}}{n}
\label{13}
\end{equation} \\
$Theorem$ $2$: Convergence rate of a ring network for even number of nodes is expressed as
\begin{equation}
R = \frac{{2 - 2a^2  - 2\cos \frac{{2\pi }}{n} + 2a^2 \cos \frac{{2\pi }}{n}}}{{3 - a^2  + ( - 1 + a^2 )\cos \frac{{2\pi }}{n}}}
\label{14}
\end{equation}
$Proof$ $2$: 
For $n$=even, $\lambda _1 (L)$ is the second smallest value of Laplacian matrix and $\lambda _{\frac{n}{2}} (L)$ is the largest of a Laplacian matrix.\\
Thus, rewrite the (\ref{8}) as
\begin{equation}
\left| {1 - h\lambda _1 (L)} \right| = \left| {1 - h\lambda _{\frac{n}{2}} (L)} \right|
\label{15}
\end{equation} \\
Substituting the expressions of $\lambda _1 (L)$ and $\lambda _{\frac{n}{2}} (L)$ in (\ref{15}), gives
\begin{equation}
h = \frac{{2 + 2\cos \frac{{2\pi }}{n}}}{{3 - \cos ^2 \frac{{2\pi }}{n} + 2\cos \frac{{2\pi }}{n} - a^2 \sin ^2 \frac{{2\pi }}{n}}}.
\label{16}
\end{equation}
Thus, convergence factor ($\gamma$) is expressed as
\begin{equation}
\gamma=\left |1-h \left(1-\cos \left(\frac{2 \pi }{n}\right)+i a \sin \left(\frac{2 \pi }{n}\right)\right)  \right |
\label{17}
\end{equation}
Substituting the $\gamma$ in (\ref{9}) proves the $Theorem$ $2$. \\
$Theorem$ $3$: Convergence rate of a ring network for odd number of nodes is expressed as
\begin{equation}
\resizebox{0.9\hsize} {!} {$R = 1 - \frac{{\sqrt {2 + 4a^2  + 2a^4  - 2( - 1 + a^4 )\cos \frac{\pi }{n} + ( - 1 + a^2 )^2 \cos \frac{{2\pi }}{n} + 2\cos \frac{{3\pi }}{n} - 2a^4 \cos \frac{{3\pi }}{n} + \cos \frac{{4\pi }}{n} - 2a^2 \cos \frac{{4\pi }}{n} + a^4 \cos \frac{{4\pi }}{n}} }}{{\sqrt 2 \left( {2 - \left( { - 1 + a^2 } \right)\cos \frac{\pi }{n} + \left( { - 1 + a^2 } \right)\cos \frac{{2\pi }}{n}} \right)}}$}
\label{18}
\end{equation}
$Proof$ $3$:
For $n$=odd, $\lambda _1 (L)$ is the second smallest eigenvalue of Laplacian matrix and $\lambda _{\frac{n-1}{2}} (L)$ is the largest eigenvalue of a Laplacian matrix.\\
Thus, rewrite the (\ref{8}) as
\begin{equation}
\left| {1 - h\lambda _1 (L)} \right| = \left| {1 -h\lambda _{\frac{{n - 1}}{2}} (L)} \right|
\label{19}
\end{equation}
Substituting the $\lambda _1 (L) $ and $\lambda _{\frac{{n - 1}}{2}} (L)$ expressions in (\ref{19}) results in 
\begin{equation}
\resizebox{0.9\hsize} {!} {$\left| {1 - h\left( {1 - \cos \frac{{2\pi }}{n} + ia\sin \frac{{2\pi }}{n}} \right)} \right| = \left| {1 - h\left( {1 + \cos \frac{\pi }{n} + ia\sin \frac{\pi }{n}} \right)} \right|$}
\label{20}
\end{equation}
Thus, we obtain 
\begin{equation}
\resizebox{0.9\hsize} {!} {$h = \frac{{2\left( {\cos \frac{\pi }{n} + \cos \frac{{2\pi }}{n}} \right)}}{{ - \cos ^2 \frac{{2\pi }}{n} + 2\cos \frac{{2\pi }}{n} - a^2 \sin ^2 \frac{{2\pi }}{n} + \cos ^2 \frac{\pi }{n} + a^2 \sin ^2 \frac{\pi }{n} + 2\cos \frac{\pi }{n}}}$}
\label{21}
\end{equation}
Finally, we get $\gamma$ as
\begin{equation}
\gamma=\left |1-h \left(1-\cos \left(\frac{2 \pi }{n}\right)+i a \sin \left(\frac{2 \pi }{n}\right)\right)  \right |
\label{22}
\end{equation}
Substituting the (\ref{22}) value in (\ref{9}) proves the $Theorem$ $3$.
\section{Explicit Formulas of Convergence Rate for Torus Networks}
Torus network with asymmetric links is as shown in the Fig. 2. In this section, we derive the explicit expressions of convergence rate for a torus network and $m$-dimensional torus networks.\\
$Theorem$ $4$: The eigenvalue of a torus network is expressed as
\begin{equation}
\resizebox{1.0\hsize} {!} {$\lambda _{j_1 ,j_2 } (L) = 2 - \cos \frac{{2\pi j_1 }}{{k_1 }} - \cos \frac{{2\pi j_2 }}{{k_2 }} + ia\left( {\sin \frac{{2\pi j_1 }}{{k_1 }} + \sin \frac{{2\pi j_2 }}{{k_2 }}} \right)$}
\label{23}
\end{equation}
$Proof$ $4$:
Cartesian product of the two ring networks results in torus network. The eigenvalue of a torus network will be the addition of eigenvalues of the corresponding ring networks \cite{load}. 
\begin{equation}
\lambda _{j_1 ,j_2 } (L) = \lambda _{j_1 } (L)+\lambda _{j_2 } (L)
\label{24}
\end{equation}
Here, we assume that the torus is formed by two ring networks with $k_1$ and $k_{2}$ nodes respectively. Then  $(j_1+1)^{th}$ eigenvalue of a Laplacian matrix for a ring network can be expressed as
\begin{equation}
\lambda _{j_1 } (L) = 1 - \cos \frac{{2\pi j_1 }}{{k_1 }} + ia\sin \frac{{2\pi j_1 }}{{k_1 }}
\label{25}
\end{equation}
Similarly, $(j_2+1)^{th}$ eigenvalue of a Laplacian matrix for ring network is expressed as
\begin{equation}
\lambda _{j_2 } (L) = 1 - \cos \frac{{2\pi j_2 }}{{k_2 }} + ia\sin \frac{{2\pi j_2 }}{{k_2 }}
\label{26}
\end{equation}
Finally, using (\ref{24}), (\ref{25}), and (\ref{26}) we obtain(\ref{23}). \\
$Theorem$ $5$: Convergence rate of a torus network for $k_1=even$ and $k_2=even$ is expressed as
\begin{equation}
\frac{a^2 \sin ^2\left(\frac{2 \pi }{k_2}\right)+\cos ^2\left(\frac{2 \pi }{k_2}\right)+6 \cos \left(\frac{2 \pi }{k_2}\right)+9}{a^2 \sin ^2\left(\frac{2 \pi }{k_2}\right)+\cos ^2\left(\frac{2 \pi }{k_2}\right)-2 \cos \left(\frac{2 \pi }{k_2}\right)-15}
\label{27}
\end{equation}
$Proof$ $5$: For $k_1=even$ and $k_2=even$ , $\lambda _{1,0} (L)$ is the second smallest eigenvalue and $\lambda _{\frac{{k_1 }}{2},\frac{{k_2 }}{2}} (L)$ is the largest eigenvalue of a Laplacian matrix. Thus, (\ref{8}) can be rewritten as
\begin{equation}
\left| {1 - h\lambda _{1,0} (L)} \right| = \left| {1 - h\lambda _{\frac{{k_1 }}{2},\frac{{k_2 }}{2}} (L)} \right|
\label{28}
\end{equation}
Substituting the $\lambda _{1,0} (L)$ and $\lambda _{\frac{{k_1 }}{2},\frac{{k_2 }}{2}} (L)$ in (\ref{28}) results in
\begin{equation}
\left| {1 - h\left( {1 - \cos \frac{{2\pi }}{{k_2 }} + ia\sin \frac{{2\pi }}{{k_2 }}} \right)} \right| = \left| {1 - 4h} \right|
\label{29}
\end{equation}
Thus, we obtain
\begin{equation}
h = \frac{{6 + 2\cos \frac{{2\pi }}{{k_2 }}}}{{15 - \cos ^2 \frac{{2\pi }}{{k_2 }} + 2\cos \frac{{2\pi }}{{k_2 }} - a^2 \sin ^2 \frac{{2\pi }}{{k_2 }}}}
\label{30}
\end{equation}
Finally, we obtain the convergence parameter $\gamma$ as
\begin{equation}
\gamma=\left| {1 - h\left( {1 - \cos \frac{{2\pi }}{{k_2 }} + ia\sin \frac{{2\pi }}{{k_2 }}} \right)} \right|
\label{31}
\end{equation}
Thus, substituting (\ref{31}) in (\ref{9}) proves the Theorem 5. \\
$Theorem$ $6$: Convergence rate of a torus network for $k_1=odd$ and $k_2=odd$ is expressed as 
\begin{equation}
R = \sqrt {\frac{{a^2 p_1^2 \sin ^2 \frac{{2\pi }}{{k_2 }}}}{{q_1^2 }} + \left( {1 - \frac{{p_1 \sin ^2 \frac{\pi }{{k_2 }}}}{{q_1 }}} \right)^2 }
\label{32}
\end{equation}
where 
\begin{equation*}
\begin{aligned}
p_1 ={} & 4 \left(2 \cos \left(\frac{\pi }{k_1}\right)+\cos \left(\frac{\pi }{k_2}\right)+\cos \left(\frac{2 \pi }{k_2}\right)+1\right),
\end{aligned}
\end{equation*}

\begin{equation*}
\begin{aligned}
q_1 ={} & -a^2 \sin ^2\left(\frac{2 \pi }{k_2}\right)+a^2 \left(\sin \left(\frac{\pi }{k_1}\right)+\sin \left(\frac{\pi }{k_2}\right)\right){}^2 \\
      & {}-\cos ^2\left(\frac{2 \pi }{k_2}\right)+\left(2 \cos \left(\frac{\pi }{k_1}\right)+\cos \left(\frac{\pi
   }{k_2}\right)\right){}^2\\
   &+8 \cos \left(\frac{\pi }{k_1}\right)+4 \cos \left(\frac{\pi }{k_2}\right)+2 \cos \left(\frac{2 \pi }{k_2}\right)+3
\end{aligned}
\end{equation*}

$Proof$ $6$: For $k_1=odd$ and $k_2=odd$, $\lambda _{1,0} (L)$ is the second smallest eigenvalue and $\lambda _{\frac{{k_1-1 }}{2},\frac{{k_2-1}}{2}} (L)$ is the largest eigenvalue of a Laplacian matrix. Thus rewrite (\ref{8}) as
\begin{equation}
\left| {1 - h\lambda _{1,0} (L)} \right| = \left| {1 - h\lambda _{\frac{{k_1-1 }}{2},\frac{{k_2-1}}{2}} (L)} \right|
\label{33}
\end{equation}
Substitute the expressions of $\lambda _{1,0} (L)$ and $\lambda _{\frac{{k_1-1 }}{2},\frac{{k_2-1}}{2}} (L)$ in (\ref{33}) results in
\begin{equation}
\resizebox{1.1\hsize} {!} {$\left| {1 - h\left( {1 - \cos \frac{{2\pi }}{{k_2 }} + ia\sin \frac{{2\pi }}{{k_2 }}} \right)} \right| = \left| {1 - h\left( {2 - \cos \frac{{\pi (k_1  - 1)}}{{k_1 }} - \cos \frac{{\pi (k_2  - 1)}}{{k_2 }} + ia\left( {\sin \frac{{\pi (k_1  - 1)}}{{k_1 }} + \sin \frac{{\pi (k_2  - 1)}}{{k_2 }}} \right)} \right)} \right|$}
\label{34}
\end{equation}
Thus, we get
\begin{equation}
\resizebox{1.1\hsize} {!} {$h=\frac{-2 \cos \left(\frac{2 \pi }{k_2}\right)+2 \left(2 \cos \left(\frac{\pi  \left(k_1-1\right)}{k_1}\right)+\cos \left(\frac{\pi  \left(k_2-1\right)}{k_2}\right)\right)-2}{0.16 \sin ^2\left(\frac{2 \pi }{k_2}\right)-0.16 \left(\sin \left(\frac{\pi 
   \left(k_1-1\right)}{k_1}\right)+\sin \left(\frac{\pi  \left(k_2-1\right)}{k_2}\right)\right){}^2+\cos ^2\left(\frac{2 \pi }{k_2}\right)-2 \cos \left(\frac{2 \pi }{k_2}\right)-\left(2 \cos \left(\frac{\pi  \left(k_1-1\right)}{k_1}\right)+\cos
   \left(\frac{\pi  \left(k_2-1\right)}{k_2}\right)\right){}^2+4 \left(2 \cos \left(\frac{\pi  \left(k_1-1\right)}{k_1}\right)+\cos \left(\frac{\pi  \left(k_2-1\right)}{k_2}\right)\right)-3}$}
\label{35}
\end{equation}
Finally, we obtain the convergence parameter $\gamma$ as
\begin{equation}
\gamma=\left| {1 - h\left( {1 - \cos \frac{{2\pi }}{{k_2 }} + ia\sin \frac{{2\pi }}{{k_2 }}} \right)} \right|
\label{36}
\end{equation}
Substituting  the (\ref{36}) in (\ref{9}) results in (\ref{32}). \\
$Theorem$ $7$: The eigenvalue of a $m$-dimensional torus network is expressed as
\begin{equation}
\lambda _{j_1 ,j_2 .....j_m } (L) = m - \sum\limits_{l = 1}^m {\cos \frac{{2\pi j_l }}{{k_l }}}  + ia\left( {\sum\limits_{l = 1}^m {\sin \frac{{2\pi j_l }}{{k_l }}} } \right)
\label{37}
\end{equation}
$Proof$ $7$: Cartesian product of `\textit{m}' ring networks results in $m$-dimenional torus network. The eigenvalue of a torus network will be the addition of eigenvalues of corresponding $m$ ring networks \cite{load}. 
\begin{equation}
\lambda _{j_1 ,j_2 .....j_m } (L)= \lambda _{j_1 } (L)+\lambda _{j_2 }(L)+....+\lambda _{j_1 } (L)+\lambda _{j_m}(L)
\label{38}
\end{equation}
Here, we assume that the torus is formed by cartesian product of '\textit{m}' ring networks with $k_m$ nodes, $m = 1,2....$. Then  $(j_1+1)^{th}$ eigenvalue of a Laplacian matrix  for ring network with $k_1$ nodes can be expressed as
\begin{equation}
\lambda _{j_1 } (L) = 1 - \cos \frac{{2\pi j_1 }}{{k_1 }} + ia\sin \frac{{2\pi j_1 }}{{k_1 }}
\label{39}
\end{equation}
The $(j_2+1)^{th}$ eigenvalue of a Laplacian matrix for ring network with $k_2$ nodes is expressed as
\begin{equation}
\lambda _{j_2 } (L) = 1 - \cos \frac{{2\pi j_2 }}{{k_2 }} + ia\sin \frac{{2\pi j_2 }}{{k_2 }}
\label{40}
\end{equation}
Similarly, $(j_m+1)^{th}$ eigenvalue of a Laplacian matrix for ring network with $k_m$ nodes is expressed as
\begin{equation}
\lambda _{j_m } (L) = 1 - \cos \frac{{2\pi j_m }}{{k_m }} + ia\sin \frac{{2\pi j_m }}{{k_m }}
\label{41}
\end{equation}
Using (\ref{38}), (\ref{39}), (\ref{40}), and (\ref{41}) we can write the eigenvalue of a $m$-dimensional torus network as (\ref{37}). \\
$Theorem$ $8$: Convergence rate of a $m$-dimensional torus network for $k_1=k_2=...k_m=even$ is expressed as
\begin{equation}
\resizebox{0.9\hsize} {!} {$R =\frac{a^2 \sin ^2\left(\frac{2 \pi }{k_1}\right)+(4 m-2) \cos \left(\frac{2 \pi }{k_1}\right)+\cos ^2\left(\frac{2 \pi }{k_1}\right)+(1-2 m)^2}{a^2 \sin ^2\left(\frac{2 \pi }{k_1}\right)+\cos ^2\left(\frac{2 \pi }{k_1}\right)-2 \cos \left(\frac{2 \pi
   }{k_1}\right)-4 m^2+1}$}
   \label{42}
\end{equation}
$Proof$ $8$: For $k_1  = k_2  = ..... = k_m =even $, second smallest eigenvalue of a Laplacian matrix is $\lambda _{1,0,0.....,0} (L)$ and largest eigenvalue of Laplacian matrix is $\lambda _{\frac{{k_1 }}{2},\frac{{k_2 }}{2},.....\frac{{k_n }}{2}} (L)$. Thus, rewrite (\ref{9}) as
\begin{equation}
\left| {1 - h\lambda _{1,0,0.....,0} (L)} \right| = \left| {1 - h\lambda _{\frac{{k_1 }}{2},\frac{{k_2 }}{2},.....\frac{{k_n }}{2}} (L)} \right|
\label{43}
\end{equation}
Substituting the $\lambda _{1,0,0.....,0} (L)$ and $\lambda _{\frac{{k_1 }}{2},\frac{{k_2 }}{2},.....\frac{{k_n }}{2}} (L)$ in (\ref{43}) results in
\begin{equation}
\left| {1 - h\left( {1 - \cos \frac{{2\pi }}{{k_1 }} + ia\sin \frac{{2\pi }}{{k_1 }}} \right)} \right| = \left| {1 - 2mh} \right|
\label{44}
\end{equation}
Thus, we obtain 
\begin{equation}
h = \frac{{2 - 2\cos \frac{{2\pi }}{{k_1 }} - 4m}}{{1 - 4m^2  + \cos ^2 \frac{{2\pi }}{{k_1 }} - 2\cos \frac{{2\pi }}{{k_1 }} + a^2 \sin ^2 \frac{{2\pi }}{{k_1 }}}}
\label{45}
\end{equation}
Substituting the $h$ in $\left| {1 - h\left( {1 - \cos \frac{{2\pi }}{{k_1 }} + ia\sin \frac{{2\pi }}{{k_1 }}} \right)} \right|$ gives $\gamma$.
Finally, substituting $\gamma$ value in (\ref{9}) results in (\ref{42}). \\
\textbf{Note}: We are unable to give the expression of Convergence rate of a $m$-dimensional torus network for $k_1=k_2=...k_m=odd$, because the expression is too long and unable to fit in this format. \\
$Theorem$ $9$: Consensus parameter of a $m$-dimensional torus network for $k_1=k_2=...k_m=odd$ is expressed as\\
\begin{equation}
\resizebox{1.0\hsize} {!} {$h = \frac{{2 - 2\cos \frac{{2\pi }}{{k_1 }} - 4m}}{{1 + \cos ^2 \frac{{2\pi }}{{k_1 }} - 2\cos \frac{{2\pi }}{{k_1 }} + a^2 \sin ^2 \frac{{2\pi }}{{k_1 }} - m^2  - \left( {\sum\limits_{i = 1}^m {\cos \frac{{\pi \left( {k_i  - 1} \right)}}{{k_i }}} } \right)^2  - a^2 \left( {\sum\limits_{i = 1}^m {\sin \frac{{\pi \left( {k_i  - 1} \right)}}{{k_i }}} } \right)^2 }}$}
\label{46}
\end{equation}
$Proof$ $9$:
For $n$=odd, second smallest eigenvalue of a Laplacian matrix is $\lambda _{1,0,0.....,0} (L)$ and largest eigenvalue of Laplacian matrix is $\lambda _{\frac{{k_1-1}}{2},\frac{{k_2-1}}{2},.....\frac{{k_n-1}}{2}} (L)$. Thus, (\ref{8}) can be rewritten as
\begin{equation}
\left| {1 - h\lambda _{1,0,0.....,0} (L)} \right| = \left| {1 - h\lambda _{\frac{{k_1  - 1}}{2},\frac{{k_2  - 1}}{2},.....\frac{{k_n  - 1}}{2}} (L)} \right|
\label{47}
\end{equation}
Substitute the $\lambda _{1,0,0.....,0} (L)$ and $\lambda _{\frac{{k_1-1}}{2},\frac{{k_2-1}}{2},.....\frac{{k_n-1}}{2}} (L)$ in (\ref{46}) results in
\begin{equation}
\resizebox{1.0\hsize} {!} {$\left| {1 - h\left( {1 - \cos \frac{{2\pi }}{{k_1 }} + ia\sin \frac{{2\pi }}{{k_1 }}} \right)} \right| = \left| {1 - h\left( {m - \sum\limits_{i = 1}^m {\cos \frac{{\pi \left( {k_i  - 1} \right)}}{{k_i }} + ia\sum\limits_{i = 1}^m {\sin \frac{{\pi \left( {k_i  - 1} \right)}}{{k_i }}} } } \right)} \right|$}
\label{48}
\end{equation}
Simplifying (\ref{48}) further results in (\ref{46}).
\section{Explicit Formulas of Convergence Rate for $r$-Nearest Neighbor Networks}
In this section, we derive the explicit expressions of convergence rate for $r$-nearest neighbor networks. In this network, nodes in the distance \textit{r} gets connected. The variable \textit{r} models the node’s transmission radius or node overhead in WSNs. \\
$Theorem$ $10$: The $(j+1)^{th}$ eigenvalue of a $r$-nearest neighbor ring network is expressed as
\begin{equation}
\lambda _j (L) = r - \sum\limits_{k = 1}^r {\cos \frac{{2\pi jk}}{n}}  + ia\sum\limits_{k = 1}^r {\sin \frac{{2\pi jk}}{n}} 
\label{49}
\end{equation}
$Proof$ $10$: Laplacian matrix of a $r$-nearest neighbor ring network with $\textit{n}$ can be written as
\begin{equation}
\resizebox{0.9\hsize} {!} {$L = circ\{ r\,\,\underbrace {\frac{{ - 1 + a}}{2}\,\,\frac{{ - 1 + a}}{2}\,\,.....\frac{{ - 1 + a}}{2}}_{r\,terms}.....\underbrace {\frac{{ - 1 - a}}{2}\,\,\frac{{ - 1 - a}}{2}....\frac{{ - 1 - a}}{2}}_{r\,\,terms}\}$}
\label{50} 
\end{equation}
Using (\ref{10}) and (\ref{50}), we obtain (\ref{49}). \\
$Theorem$ $11$: Convergence Rate of a $r$-nearest neighbor ring network for $n=even$ is expressed as
\begin{equation}
R = 1 - \sqrt {\left( {\frac{{p_2 q_2 }}{{s_2 }}} \right)^2  + \frac{{r_2 q_2 }}{{s_2 }} + 1} 
\label{51}
\end{equation}, where
\begin{equation*}
\begin{aligned}
p_2 ={} & \frac{{a\sin \frac{\pi }{n}\cos \frac{{(2r + 1)\pi }}{n} - 0.5a\sin \frac{{2\pi }}{n}}}{{\cos \frac{{2\pi }}{n} - 1}}, \\
 q_2 ={} & \frac{{\sin \frac{{(2r + 1)\pi }}{n}}}{{\sin \frac{\pi }{n}}} - \cos \pi r ,\\
 r_2 ={}  & \frac{{\sin \frac{\pi }{n}\sin \frac{{(2r + 1)\pi }}{n}}}{{\cos \frac{{2\pi }}{n} - 1}} + r + 0.5,
\end{aligned}
\end{equation*}
 and
 
\begin{equation*}
\begin{aligned}
s_2 ={} & \frac{0.25 a^2 \left(2 \sin \left(\frac{\pi }{n}\right) \cos \left(\frac{2 \pi  r+\pi }{n}\right)-\sin \left(\frac{2 \pi }{n}\right)\right)^2}{\left(\cos \left(\frac{2 \pi }{n}\right)-1\right)^2} \\
      & +\frac{\sin ^2\left(\frac{\pi }{n}\right) \sin
   ^2\left(\frac{2 \pi  r+\pi }{n}\right)}{\left(\cos \left(\frac{2 \pi }{n}\right)-1\right)^2}\\
   &+(r+0.5) \left(\cos (\pi  r)-\csc \left(\frac{\pi }{n}\right) \sin \left(\frac{2 \pi  r+\pi }{n}\right)\right)\\
   &-0.25 \cos ^2(\pi  r). 
\end{aligned}
\end{equation*}

$Proof$ $11$: 
For $n$=even, $\lambda _1 (L)$ is the second smallest value of Laplacian matrix and $\lambda _{\frac{n}{2}} (L)$ is the largest eigenvalue of a Laplacian matrix. Thus, rewrite (\ref{9}) as
\begin{equation}
\left| {1 - h\lambda _1 (L)} \right| = \left| {1 - h\lambda _{\frac{n}{2}} (L)} \right|
\label{52}
\end{equation} \\
Substituting the $\lambda _1 (L)$ and $\lambda _{\frac{n}{2}} (L)$ in (\ref{52}), results in
\begin{equation}
\resizebox{1.0\hsize} {!} {$\left| {1 - h\left( {r + 0.5 - \frac{{\sin \frac{{(2r + 1)\pi }}{n}}}{{2\sin \frac{\pi }{n}}} + \frac{{ia}}{2}\left( {\cot \frac{\pi }{n} - \frac{{\cos \frac{{(2r + 1)\pi }}{n}}}{{\sin \frac{\pi }{n}}}} \right)} \right)} \right| = \left| {1 - h\left( {r + 0.5 - 0.5\cos \pi r} \right)} \right|$}
\label{53}
\end{equation}
Thus, we get
\begin{equation}
\resizebox{0.9\hsize} {!} {$h = \frac{{\cos \pi r - \frac{{\sin \frac{{(2r + 1)\pi }}{n}}}{{\sin \frac{\pi }{n}}}}}{{\frac{1}{4}\left( {\frac{{\sin \frac{{(2r + 1)\pi }}{n}}}{{\sin \frac{\pi }{n}}}} \right)^2  - (r + 0.5)\left( {\frac{{\sin \frac{{(2r + 1)\pi }}{n}}}{{\sin \frac{\pi }{n}}} - \cos \pi r} \right) - \frac{{\cos ^2 \pi r}}{4} + \frac{{e^2 }}{4}\left( {\cot \frac{\pi }{n} - \frac{{\cos \frac{{(2r + 1)\pi }}{n}}}{{\sin \frac{\pi }{n}}}} \right)^2 }}$}
\label{54}
\end{equation}
Thus, convergence factor is expressed as
\begin{equation}
\resizebox{0.9\hsize} {!} {$\gamma=\left| {1 - h\left( {r + 0.5 - \frac{{\sin \frac{{(2r + 1)\pi }}{n}}}{{2\sin \frac{\pi }{n}}} + \frac{{ia}}{2}\left( {\cot \frac{\pi }{n} - \frac{{\cos \frac{{(2r + 1)\pi }}{n}}}{{\sin \frac{\pi }{n}}}} \right)} \right)} \right|$}
\label{55}
\end{equation}
Substitute (\ref{55}) in (\ref{9}) proves the $Theorem$ $11$. \\
$Theorem$ $12$: Convergence Rate of a $r$-nearest neighbor ring network for $n=odd$ is expressed as
\begin{equation}
R = 1 - \sqrt {\left( {\frac{{p_3 q_3 }}{{s_3 }}} \right)^2  + \frac{{r_3 q_3 }}{{s_3 }} + 1} 
\label{56}
\end{equation},
where

\begin{equation*}
\begin{aligned}
p_3 ={} & \frac{{ - a\sin \frac{\pi }{n}\cos \frac{{(2r + 1)\pi }}{n} + 0.5a\sin \frac{{2\pi }}{n}}}{{\cos \frac{{2\pi }}{n} - 1}}, \\
 q_3 ={} & - \frac{{\sin \frac{{(2r + 1)\pi }}{n}}}{{\sin \frac{\pi }{n}}} + \frac{{\sin \frac{{\pi (n - 1)(2r + 1)}}{{2n}}}}{{\cos \frac{\pi }{{2n}}}} ,\\
 r_3 ={}  & - \frac{{\sin \frac{\pi }{n}\sin \frac{{(2r + 1)\pi }}{n}}}{{\cos \frac{{2\pi }}{n} - 1}} - r - 0.5,
\end{aligned}
\end{equation*}
and
\begin{equation*}
\begin{aligned}
s_3 ={} & -\frac{0.25 a^2 \left(2 \cos \left(\frac{\pi }{2 n}\right) \cos \left(\frac{\pi  (n-1) (2 r+1)}{2 n}\right)-\sin \left(\frac{\pi }{n}\right)\right)^2}{\left(\cos \left(\frac{\pi }{n}\right)+1\right)^2} \\
      & +\frac{0.25 a^2 \left(2 \sin \left(\frac{\pi}{n}\right) \cos \left(\frac{2 \pi  r+\pi }{n}\right)-\sin \left(\frac{2 \pi }{n}\right)\right)^2}{\left(\cos \left(\frac{2 \pi }{n}\right)-1\right)^2} \\
      & -\frac{\cos ^2\left(\frac{\pi }{2 n}\right) \sin ^2\left(\frac{\pi  (n-1) (2 r+1)}{2n}\right)}{\left(\cos \left(\frac{\pi }{n}\right)+1\right)^2}+\frac{\sin ^2\left(\frac{\pi }{n}\right) \sin ^2\left(\frac{2 \pi  r+\pi }{n}\right)}{\left(\cos \left(\frac{2 \pi }{n}\right)-1\right)^2} \\
      &+(r + 0.5)\left( {\frac{{\sin \frac{{\pi (n - 1)(2r + 1)}}{{2n}}}}{{\cos \frac{\pi }{{2n}}}} - \frac{{\sin \frac{{(2r + 1)\pi }}{n}}}{{\sin \frac{\pi }{n}}}} \right).
\end{aligned}
\end{equation*}
$Proof$ $12$:
For $n$=odd, $\lambda _1 (L)$ is the second smallest eigenvalue of Laplacian matrix and $\lambda _{\frac{n-1}{2}} (L)$ is the largest eigenvalue of a Laplacian matrix.\\
Thus, we rewrite the (\ref{8}) as
\begin{equation}
\left| {1 - h\lambda _1 (L)} \right| = \left| {1 - h\lambda _{\frac{n-1}{2}} (L)} \right|
\label{57}
\end{equation} \\
After substituting the $\lambda _1 (L)$ and $\lambda _{\frac{n-1}{2}} (L)$ expressions in (\ref{57}), we obtain
\begin{equation}
\resizebox{1.1\hsize} {!} {$\left| {1 - h\left( {r + 0.5 - \frac{{\sin \frac{{(2r + 1)\pi }}{n}}}{{2\sin \frac{\pi }{n}}} + \frac{{ia}}{2}\left( {\cot \frac{\pi }{n} - \frac{{\cos \frac{{(2r + 1)\pi }}{n}}}{{\sin \frac{\pi }{n}}}} \right)} \right)} \right| = \left| {1 - h\left( {r + 0.5 - \frac{{\sin \frac{{(2r + 1)\pi (n - 1)}}{{2n}}}}{{\sin \frac{{\pi (n - 1)}}{{2n}}}} + \frac{{ia}}{2}\left( {\cot \frac{{\pi (n - 1)}}{{2n}} - \frac{{\cos \frac{{(2r + 1)\pi (n - 1)}}{{2n}}}}{{\sin \frac{{\pi (n - 1)}}{{2n}}}}} \right)} \right)} \right|$}
\label{58}
\end{equation}
Thus, we obtain
\begin{equation}
\resizebox{1.1\hsize} {!} {$h = \frac{{\frac{{\sin \frac{{(2r + 1)\pi (n - 1)}}{{2n}}}}{{\sin \frac{{\pi (n - 1)}}{{2n}}}} - \frac{{\sin \frac{{(2r + 1)\pi }}{n}}}{{\sin \frac{\pi }{n}}}}}{{\frac{1}{4}\left( {\frac{{\sin \frac{{(2r + 1)\pi }}{n}}}{{\sin \frac{\pi }{n}}}} \right)^2  + (r + 0.5)\left( { - \frac{{\sin \frac{{(2r + 1)\pi (n - 1)}}{{2n}}}}{{\sin \frac{{\pi (n - 1)}}{{2n}}}} + \frac{{\sin \frac{{(2r + 1)\pi }}{n}}}{{\sin \frac{\pi }{n}}}} \right) - \frac{1}{4}\left( {\frac{{\sin \frac{{(2r + 1)\pi (n - 1)}}{{2n}}}}{{\sin \frac{{\pi (n - 1)}}{{2n}}}}} \right)^2  + \frac{{e^2 }}{4}\left( {\cot \frac{\pi }{n} - \frac{{\cos \frac{{(2r + 1)\pi }}{n}}}{{\sin \frac{\pi }{n}}}} \right)^2  - \frac{{e^2 }}{4}\left( {\cot \frac{{\pi (n - 1)}}{{2n}} - \frac{{\cos \frac{{(2r + 1)\pi (n - 1)}}{{2n}}}}{{\sin \frac{{\pi (n - 1)}}{{2n}}}}} \right)^2 }}$}
\label{59}
\end{equation}
Finally, convergence factor $\gamma$ is expressed as
\begin{equation}
\resizebox{1.05\hsize} {!} {$\gamma=\left| {1 - h\left( {r + 0.5 - \frac{{\sin \frac{{(2r + 1)\pi }}{n}}}{{2\sin \frac{\pi }{n}}} + \frac{{ia}}{2}\left( {\cot \frac{\pi }{n} - \frac{{\cos \frac{{(2r + 1)\pi }}{n}}}{{\sin \frac{\pi }{n}}}} \right)} \right)} \right|$}
\label{60}
\end{equation}
Substituting  the (\ref{60}) in (\ref{9}) proves the $Theorem$ $12$.
\section{Numerical Results}
In this section, we present the numerical results to investigate the effect of asymmetric link factor, network dimension, number of nodes, and node overhead on the convergence rate of the average consensus algorithm. We have used the \textit{Wolfram Mathematica} to solve the equations. Fig. 3 shows the comparison of convergence rates of asymmetric and symmetric ring networks. We have observed that the convergence rate decreases with both the number of nodes and asymmetric link weight. Fig. 4 shows the convergence rate versus $k_{1}$ and $k_{2}$. Here, convergence rate decreases with $k_{1}$ and $k_{2}$ exponentially. Fig. 5 shows the convergence rate versus asymmetric link factor for different values of $r$. We have noted that the convergence rate increases with the node overhead and decreases with asymmetric link factor. We noted that the convergence rate becomes `0' at asymmetric factor $0.8$. To understand the effect of network dimension on the convergence rate, we plotted the Fig. 6. We have observed that the convergence rate decreases with the network dimension. To compute the error introduced by the symmetric network modeling, we compute the absolute error $R_{s}-R_{a}$, where  $R_{s}$ and $R_{a}$ denote the convergence rates of symmetric and asymmetric networks respectively. Fig. 7 shows the Absolute Error versus Number of nodes. Here, the absolute error decreases with the number of nodes. Therefore, the effect of asymmetric link modeling on the convergence rate is high in small-scale networks. We have observed that the absolute error is significant for large values of asymmetric link factors.  
\begin{figure}[tbp]
\centering
\includegraphics[width=8cm,height=5.5cm]{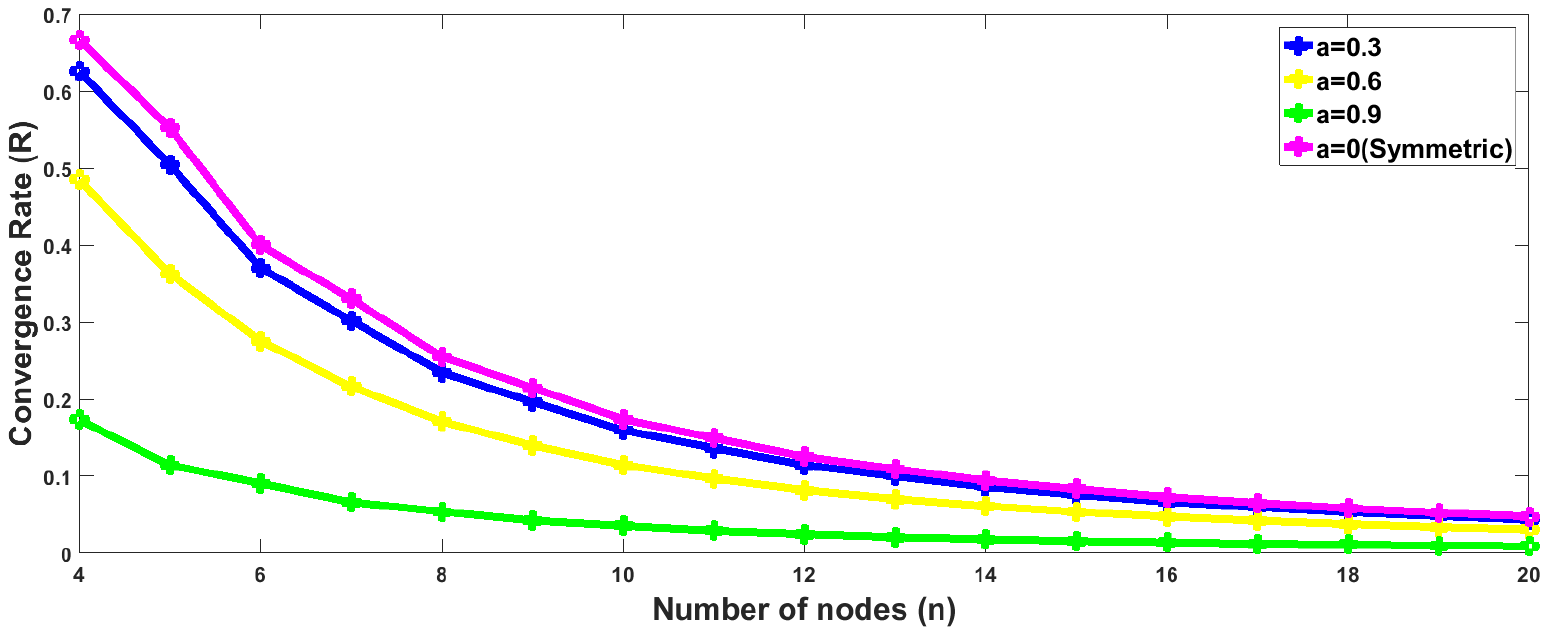}
\caption{Comparison of convergence rates in asymmetric and symmetric ring networks.}
\label{fig:verticalcell}
\end{figure}
\begin{figure}[tbp]
\centering
\includegraphics[width=6.5cm,height=5cm]{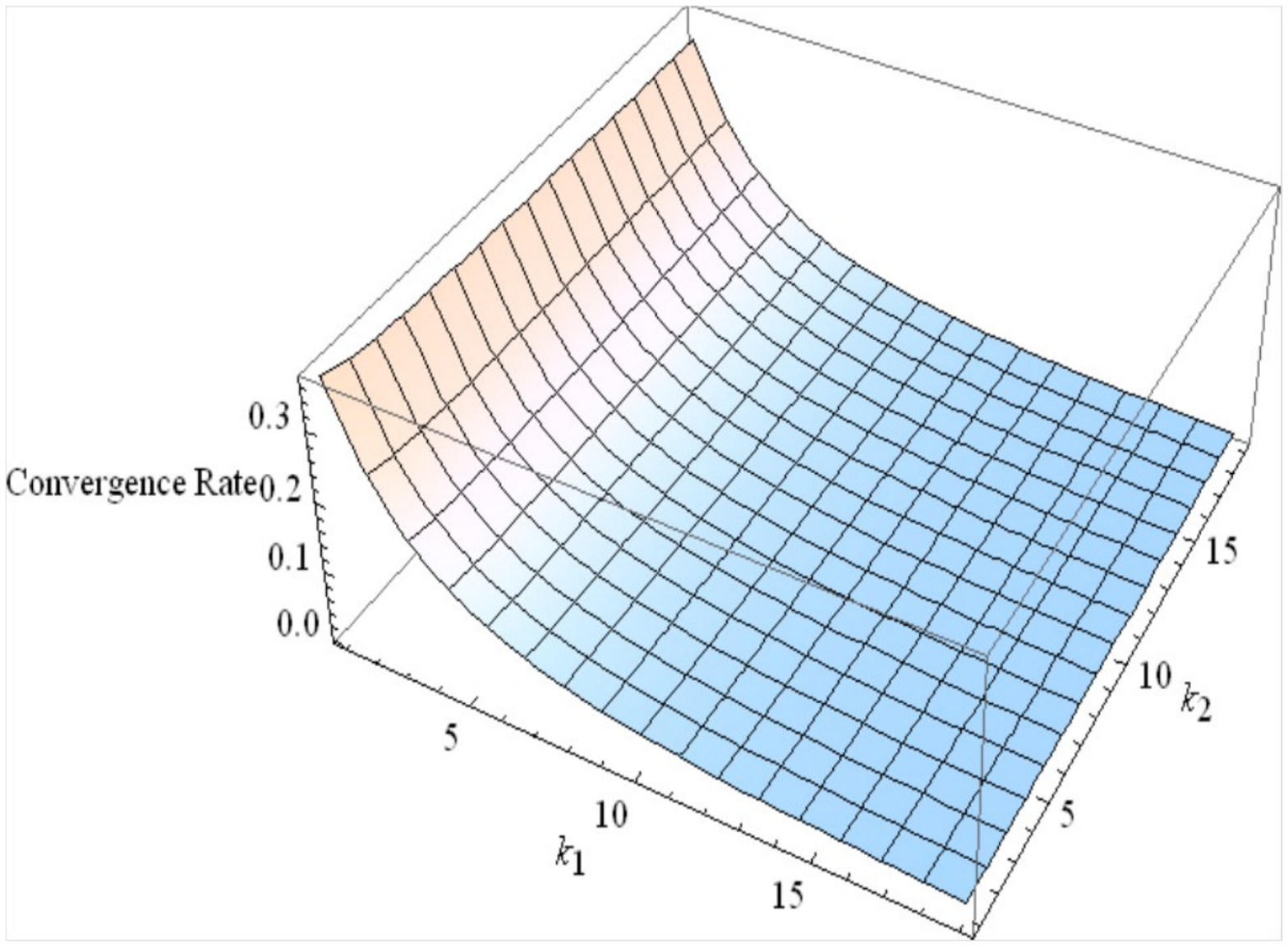}
\caption{Convergence Rate versus $k_{1}$ versus $k_{2}$ of a torus network for $n$=odd.}
\label{fig:verticalcell}
\end{figure}
\begin{figure}[tbp]
\centering
\includegraphics[width=6.5cm,height=6.5cm]{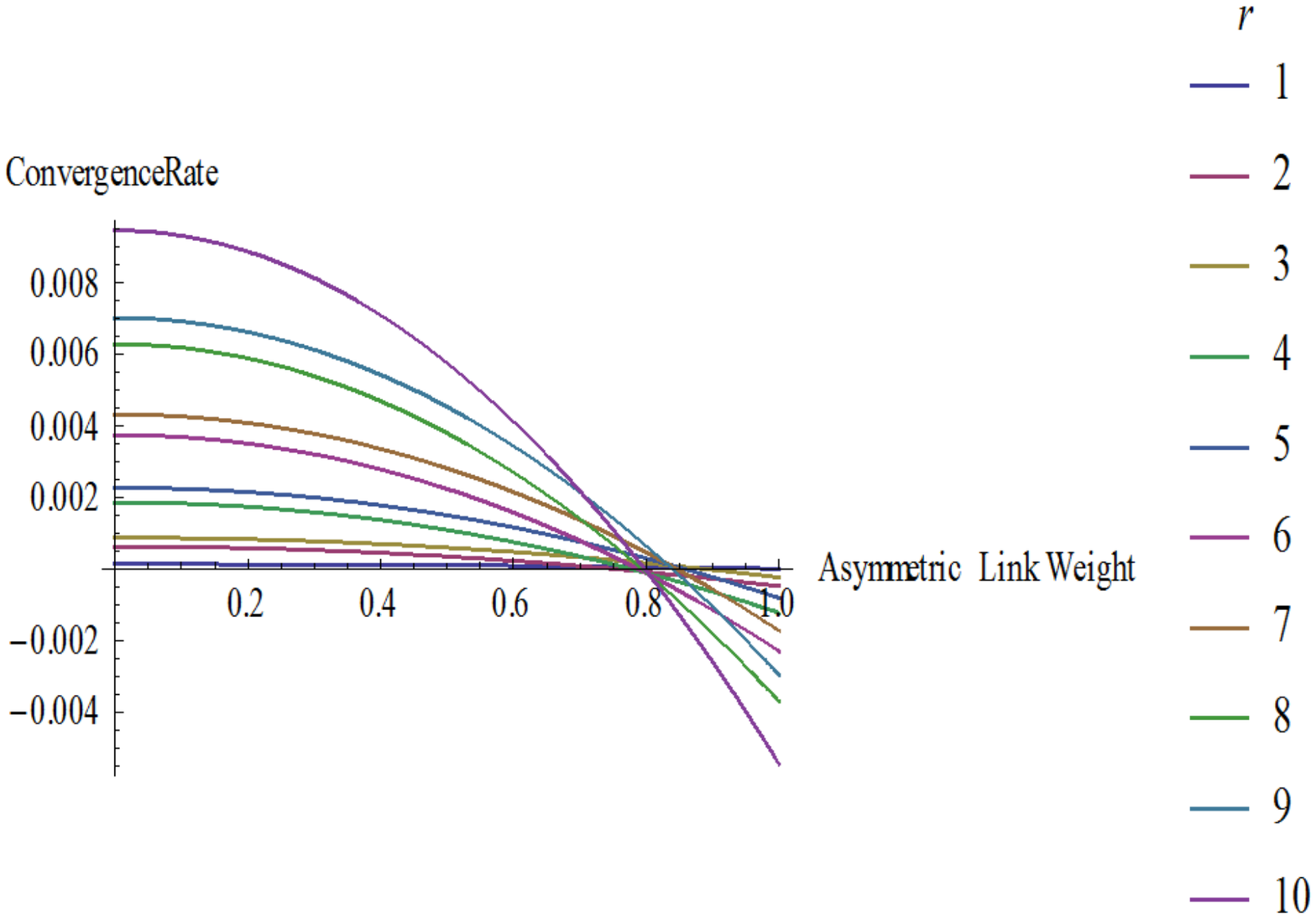}
\caption{Convergence Rate versus Asymmetric Link Weight of a $r$-nearest neighbor network for $n$=400.}
\label{fig:verticalcell}
\end{figure}
\begin{figure}[tbp]
\centering
\includegraphics[width=6.5cm,height=5cm]{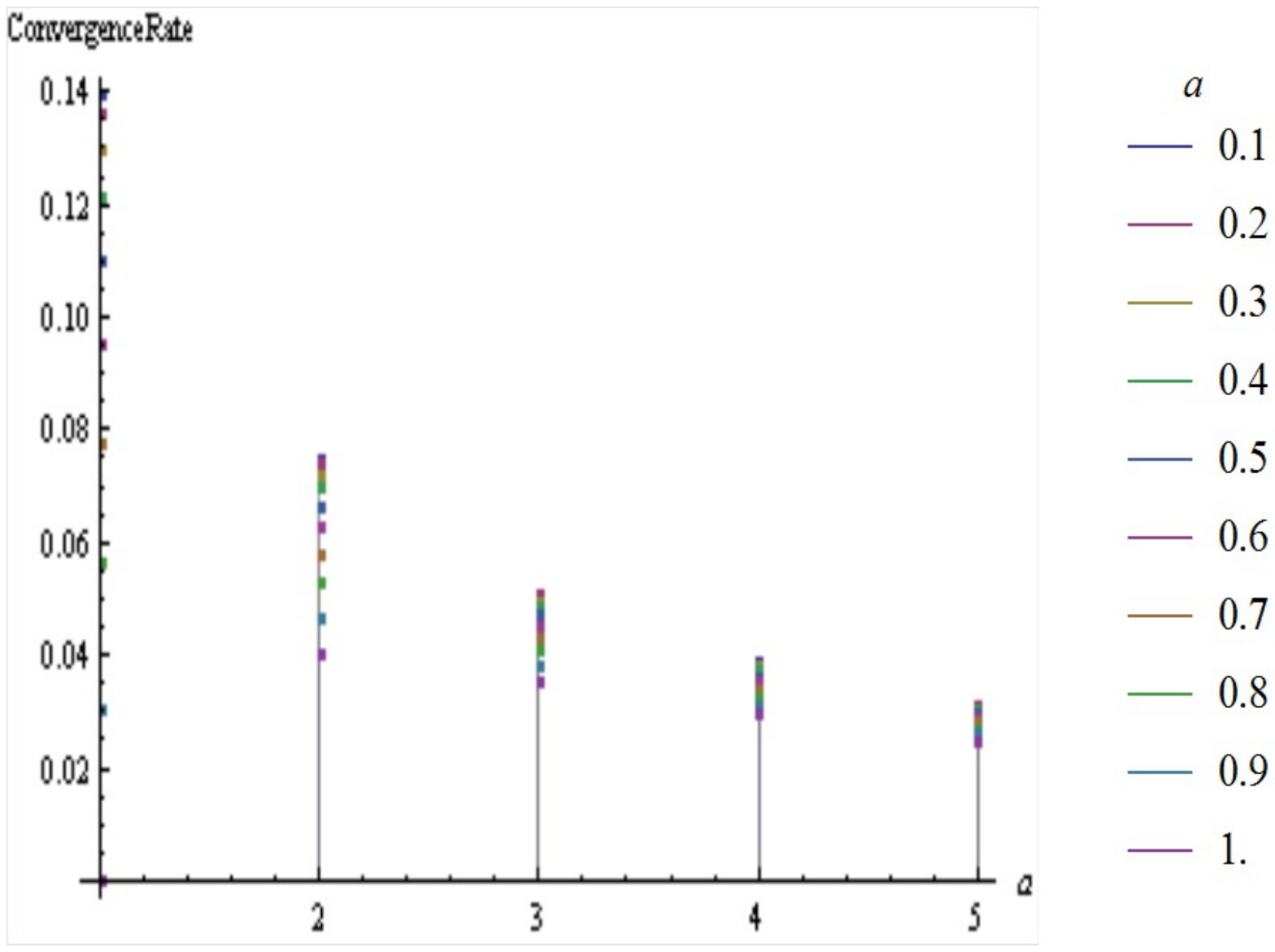}
\caption{Convergence Rate versus Network Dimension for $k_{1}=11$, $k_{2}=15$, $k_{3}=21$, $k_{4}=25$, $k_{5}=27$.}
\label{fig:verticalcell}
\end{figure}
\begin{figure}[tbp]
\centering
\includegraphics[width=6.5cm,height=5cm]{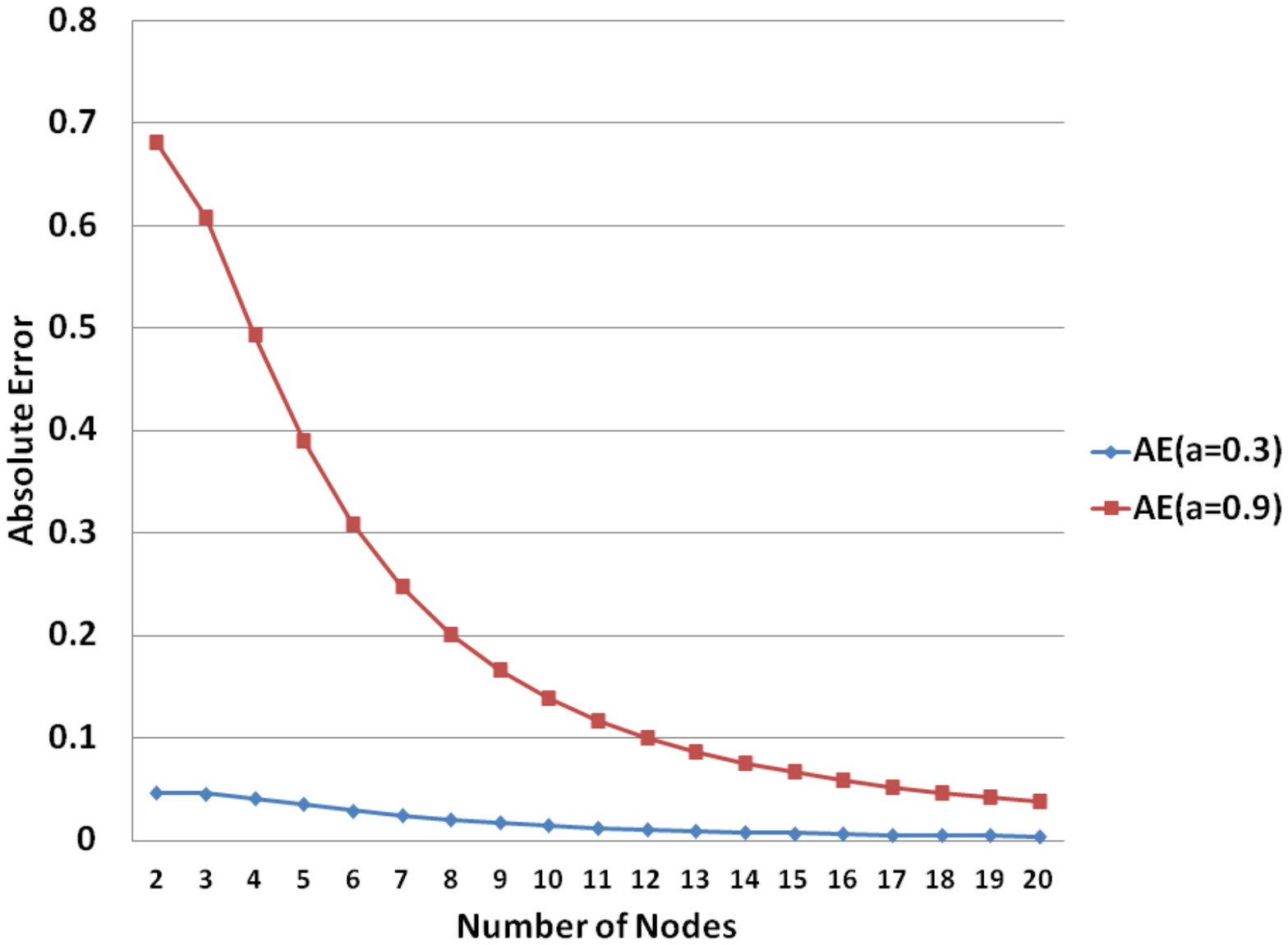}
\caption{Absolute Error versus Number of nodes for $a=0.3$, $a=0.9$.}
\label{fig:verticalcell}
\end{figure}
\section{Conclusions}
In this work, we modeled the WSN as a directed graph and derived the explicit formulas for a ring, torus, $r$-nearest neighbor ring, and $m$-dimensional torus networks. Numerical results
demonstrated that the convergence rate decreases significantly with asymmetrical link factor in small-scale WSNs. In large-scale WSNs, the effect of asymmetrical links on convergence
rate decreases with the number of nodes. Further, we studied the impact of the number of nodes, network dimension, and node overhead on the convergence rate. We have observed that the convergence rate increases with the node overhead and decreases with node dimension. However, energy consumption rises with the node overhead. Since WSNs are energy constrained networks,
it is essential to design an optimal framework to maximize the convergence rate without affecting the energy consumption. Our analysis avoids the use of sophisticated algorithms to study the convergence rate and also reduces the computational complexity drastically over the existing approaches.\ifCLASSOPTIONcaptionsoff
\newpage \fi
\bibliographystyle{IEEEtran}
\bibliography{References}


\end{document}